# Investigation on structural, electronic and magnetic properties of $Co_2FeGe$ Heusler alloy: experiment and theory

**Srimanta Mitra[a,b], Aquil Ahmad[a], Shamik Chakrabarti[c], Sajib Biswas[a] and Amal Kumar Das[a*]**

**[a] Indian Institute of Technology Kharagpur, Khargapur-721302, India**

**[b] Space Applications Centre, ISRO, Ahmedabad-380015, India**

**[c] Indian Institute of Technology Patna, Patna-800013, India**

**email: srimanta_44@sac.isro.gov.in**

**Abstract**

Experimental and computational studies were performed on $Co_2FeGe$ Heusler alloy. It was found that the alloy has very high experimental magnetic moment of 6.1 $\mu_B$/f.u., curie temperature of 1073K and very high spin-wave stiffness constant of 10.4 $nm^2$-meV, which indicates that the magnetic moment is very high and do not vary with change in temperature in the range 0-300K. The alloy strictly follows Slater-Pauling (SP) rule and the minor experimental deviation from its SP value is justified by doing full-potential density functional calculations which gives more accurate result when electron-electron correlation parameter (U) is taken into account with conventional GGA method. Effect of lattice strain and electron correlation on individual atomic moments, total magnetic moment and spin-polarization is studied in detail and can be concluded that they have a role in the deviation of the experimental results from the expected theoretical values.

## 1. Introduction

Since the discovery in 1903 by Fritz Heusler, Heusler alloys (HAs) have been proved to be potential candidates for spintronic device applications. Due to the structural similarity with the binary semiconductors [1] and tunability of properties by simply changing the numbers of valence electrons [2], theoretical research is still going on to find new Heusler alloys for specific application. Different types of HAs show various properties which are equally important for theoretical discussions and experimental applications; for example, non-magnetic full HAs with 2:1:1 stoichiometry and having number of valence electrons equal to 27 are superconductors [2][3][4], full Heusler alloys with stoichiometry 2:1:1 and half

HAs with stoichiometry 1:1:1 and having number of valence electrons equal to 24 are semiconducting in nature [5] and potential candidates for thermoelectric applications [5][6]. Half HAs are one of the most tunable class of compounds that show topological insulating behaviour [5], anomalous Hall effect [2][8][9][10] and magneto-optic phenomena [11][12][34]. Due to the presence of two magnetic sub-lattices in full Heusler alloys [13], they may allow antiferromagnetic coupling between the atomic moments which can lead to antiferromagnetism [14] or compensated ferrimagnetism [2][15]. Ni-based HAs are promising candidates for shape memory applications [16].

For spintronics device applications, highly spin-polarized materials with large magneto-resistance is required [17]; from this point of view cobalt based full HAs are of particular interest for their half-metallicity [18], high magnetic moment, high spin polarization and very high curie temperature [1][19]. They are very useful for device applications e.g. spin-voltage generators [20] and mostly for magnetic tunnel junctions [1]. For Cobalt based HAs magnetic moment increases linearly with number of valence electrons and Slater-Pauling (SP) rule is strictly followed [2][18]. For these reasons, we are motivated to study cobalt based HAs and choose $Co_2FeGe$ as our material to study in this paper. We show that this alloy shows very high magnetic moment, high curie temperature and half-metallicity which may be suitable for a magnetic tunnel junction applications [1].

In this paper, we present thorough computational and experimental studies on $Co_2FeGe$ full HA. Section.2 describes the experimental section focusing on the structural and magnetic properties of the alloy. Section.3 deals with the computational section showing the first principle density functional calculations done with the bulk alloy. The deviation of the experimental results from the expected value is justified by the computational results and is discussed in section.4.

## 2. Experimental Section

### 2. a. Methods

Polycrystalline bulk $Co_2FeGe$ alloy was prepared by repeated melting of stoichiometric mixture of 99.9% pure Co, 99.9% pure Fe and 99.999% pure Ge in an arc furnace in argon atmosphere. As the weight loss was negligible, the prepared sample was assumed to be of correct proportion and was vacuum sealed in a quartz tube, heated at 800$^{o}$C for 96 hours before further characterizations.

### 2. b. Structural characterizations

For the identification of crystallographic structure and phase, x-ray diffraction (XRD) technique was used. Figure 1. shows the simulated and experimental XRD plots for $Co_2FeGe$ HA. As it is obvious from the experimental curve, no (111) and (200) reflections, which are necessary to conclude about $L2_1$ structure, is practically observed.

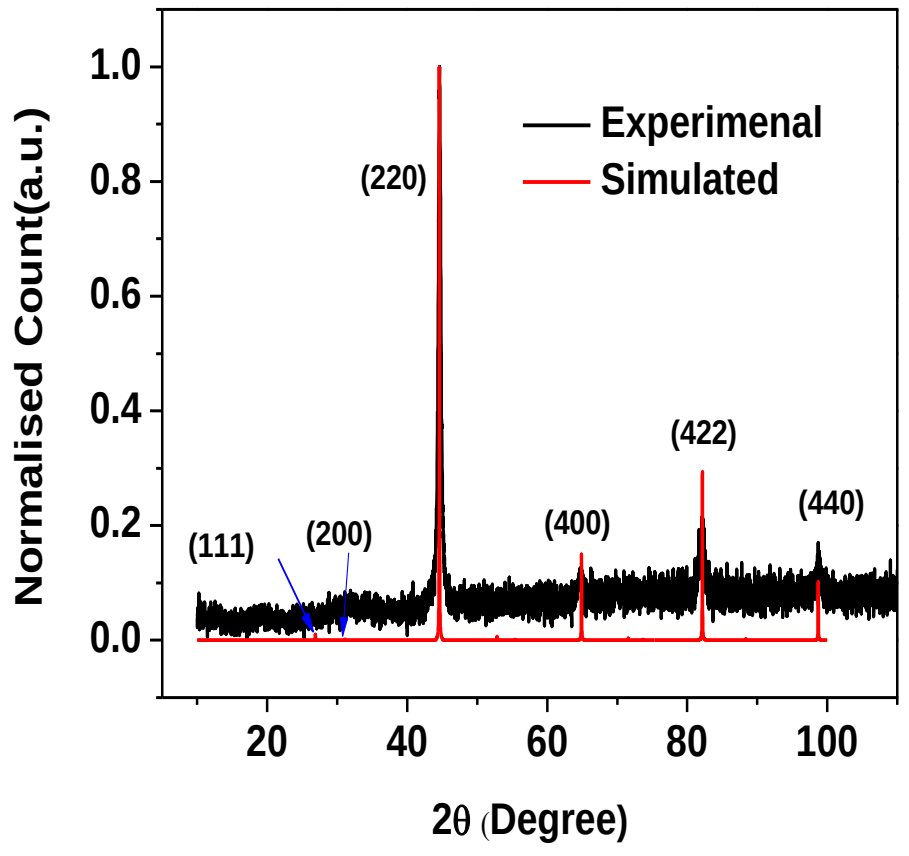

**Fig.1.** Simulated and experimental XRD patterns of $Co_2FeGe$ Heusler alloy

Apparently, it may be concluded that the structure is having A2 disorder but it is to be noted that in case of HA family having $X_2YZ$ structure where X and Y are two transition metals from same period in periodic table having nearly equal scattering factors [21], XRD is not sufficient to conclude about proper structural information as different types of disorders (if present) are not distinguishable from this technique [1] [2]. In that case extended x-ray absorption fine structure (EXAFS) is a more suitable technique for proper analysis [1] [21]. The simulated result was generated using PowderCell software. The experimental data was found with Cu $K_\alpha$ source. Cu Kα source can't give any information about disorderedness as the atomic scattering factor for Co and Fe is almost similar for Cu $K_\alpha$ source [22]. To confirm the stoichiometry of the elements in the alloy energy dispersive x-ray analysis (EDAX) technique was used. Table 1. shows the atomic weight percentage of the elements present in the alloy.

**Table 1** Atomic weight percentage of the elements in $Co_2FeGe$ HA obtained from energy dispersive x-ray analysis (EDAX) technique.

| Element | Atomic wt. % |
|---|---|
| Co | 52.53 |
| Fe | 25.46 |
| Ge | 22.19 |

## 2. c. Magnetic Characterization

Thermo-magnetic measurement was done with LAKESHORE vibrating sample magnetometer (VSM) in the temperature range 300 K to 1300 K. The sharp transition in the dm/dT vs T (m is normalized magnetization and T is temperature) graph in fig.2.b. (in-set) gives the curie temperature value of 1073K. Next we measure the magnetization as a function of magnetic field (fig. 2.a.) at different temperatures. It is noticed that the saturation magnetization is very stiff against temperature and it changes only by 0.26 $\mu_B$/f.u. in the whole temperature range of 0K to 300K.

.

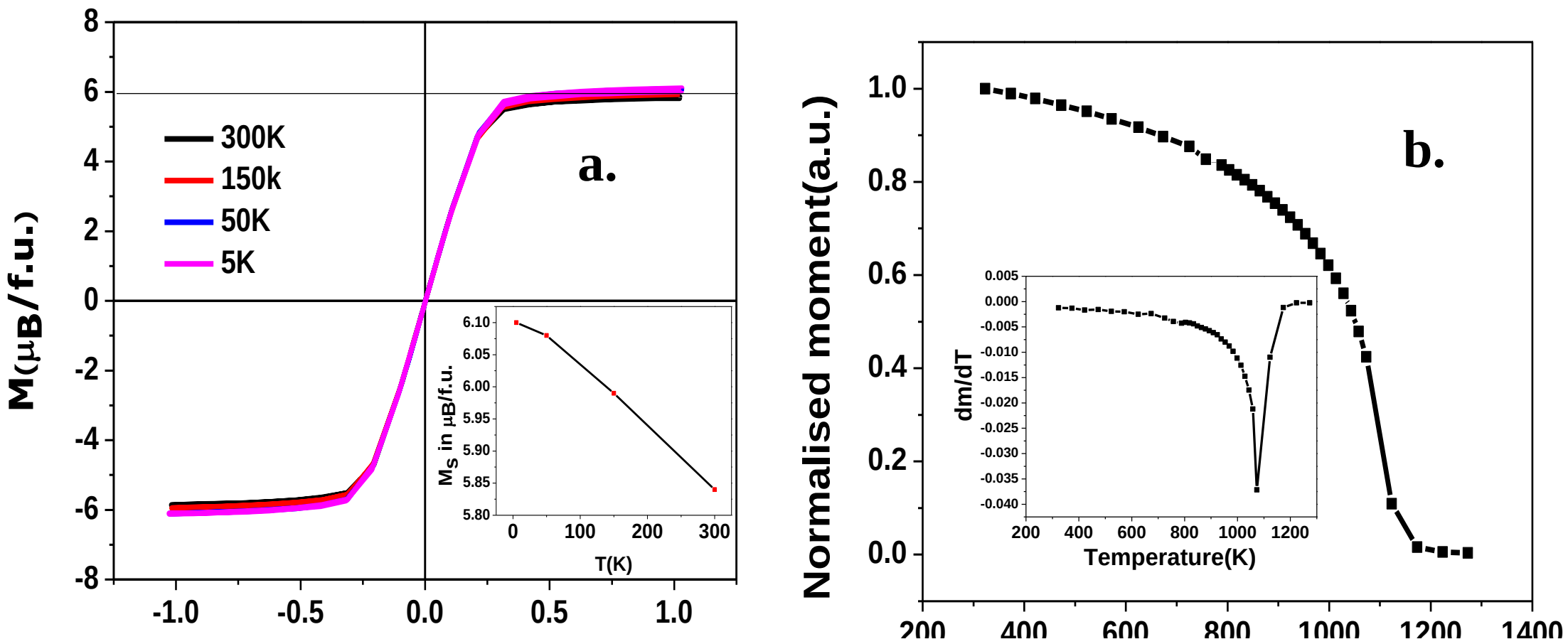

**Fig.2. a.** Magnetisation curve for $Co_2FeGe$ HA at different temperature, **b.** Dependence of normalized magnetic moment on temperatue (in-set figure shows temperature rate of change of moment with temperature marking the Curie temperature)

To get a microscopic view about this strong magnetic stiffness phenomena we focus our attention towards spin-wave theory and get acquainted with some related parameters which have a significant effect on the magnetic phase stability against temperature. Electron-electron correlation effect also has an important consequence on the temperature stability of spin-polarization in HAs [25]. An important parameter which gives information about the magnetic phase stability of the $Co_2FeGe$ HA, is the spin-wave stiffness constant (D). It describes the amount of energy of a spin-wave (neglecting dipolar interaction) and it represents the curvature in the magnon dispersion curve (E vs q) at the minimum of q=0 (Γ-point), where q is the wave vector [26]. Another parameter, called exchange constant (A) signifies the strength of exchange interaction between two spins in a ferromagnetic system [25]. In a continuous magnetization model it gives the exchange energy density (ξ). The parameters are related to each other by the following relations [25]:

$$E = \hbar\omega = (JSa^2)q^2 = Dq^2 \quad \text{(neglecting dipolar interaction)} \quad .... (1)$$

$$A = \frac{DM_S}{2g\mu_B} \quad .... (2)$$

$$\xi = \frac{E}{V} = A|\nabla \boldsymbol{m}|^2 \quad .... (3)$$

where, E is the energy of single magnon (i.e. spin-wave), ω is the angular frequency of the spin-wave, ħ is the reduced Planck constant, J is the exchange integral, a is lattice parameter, $M_s$ is saturation magnetization, g is Lande-*g*-factor, $\mu_B$ is Bohr magneton, m(=M/Ms) is reduced magnetization.

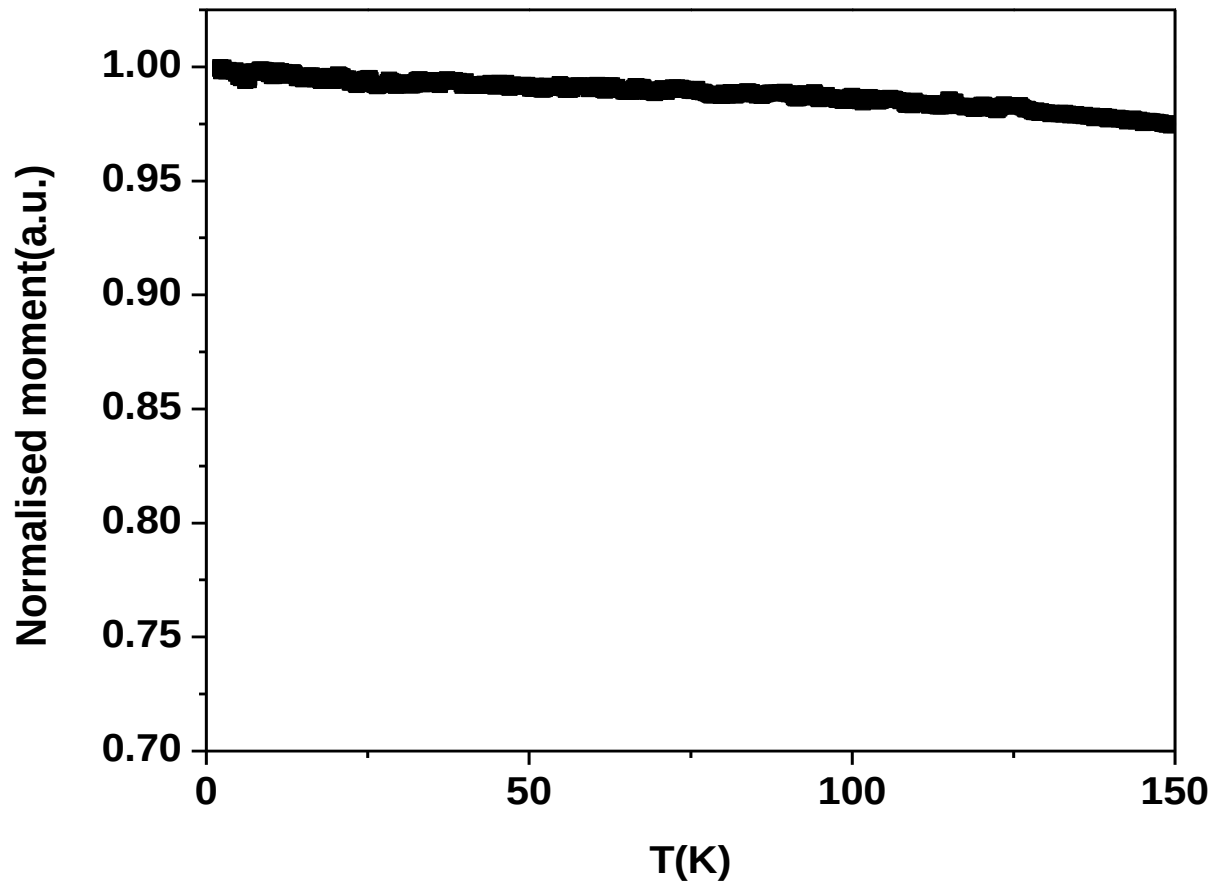

**Fig.3.** Temperature dependent saturation magnetization (normalized) curve for $Co_2FeGe$ HA

Experimentally we have measured D and A for $Co_2FeGe$ HA by the temperature variation of its saturation magnetization ($M_s$) by means of physical property measurement system (PPMS). The temperature dependence of saturation magnetization ($M_s$) is found from Bloch's law [27] [28] as:

$$M_s(T) = M_s(0K)\left(1-\gamma T^{\frac{3}{2}}\right) \text{ where}, \gamma = 2.612\frac{V_0}{\langle S\rangle}\left(\frac{k_B}{4\pi D}\right)^{\frac{3}{2}} \quad \ldots.(4)$$

# 3. Computational section

## 3. a. Methods and details

The ground state calculations were done using full-potential linearized augmented plane wave (FPLAPW) methods as implemented in the WIEN2K package [29] within the scope of generalized gradient approximation (GGA) for the exchange correlation potential as parameterized by Perdew, Burke and Ernzerhof (PBE) [30]. GGA+U approximations was done to take into account strong correlation between spin polarized electrons of transition metal elements. In FPLAPW method fully relativistic treatment was done for the core states and a scalar relativistic approximation was used for valence states [Ref]. In these calculations, the plane wave cut-off parameter determining the number of basis functions was decided by $R_{MT}K_{MAX}$ =7,where $R_{MT}$ is the smallest of all atomic muffin-tin sphere's radii and $K_{MAX}$ is the largest wave vector in the basis set. The muffin-tin radii for Co, Fe and Ge were chosen to be 2.20, 2.20 and 2.12 respectively ensuring nearly touching spheres. The calculation was done in a way that the unit cell was divided into two regions: the non-overlapping nearly touching muffin-tin spheres and interstitial regions. The magnitude of the largest vector in the charge density Fourier expansion was fixed to be $G_{max}$=12. The cut-off energy separating the valence and core states was chosen to be -6 Ry. The energy and charge convergence criterion was set respectively to be $10^{-3}$ Ry and $10^{-3}$ Coulomb during the self-consistency cycle. We have used (10 x 10 x 10) meshes i.e. 1000 k-points in the first brillouin zone. The k-space integration was done by modified tetrahedron method [31]. The effect of spin-orbit interaction is neglected in the calculations.

### 3. b. Structural properties

The $X_2YZ$ type full HAs can be found in two types of structures: $Cu_2MnAl$-type $L2_1$ Structure or $Hg_2CuTi$-type structure. Generally, the latter one is found when Z(Y) > Z(X), where X and Y are transition elements from same period in the periodic table [1]. The structural details can be found in table 2. The variation of total energy as a function of lattice volume per formula unit for different structures is plotted in fig.4 which shows that $Cu_2MnAl$-type is structurally more stable than $Hg_2CuTi$-type structure.

The variation is given by the Birch-Murnaghan's equation of state as [32] [33] [34] [35]:

$$E(V) = E_0 + \frac{9V_0B_0}{16}\left\{\left[\left(\frac{V_0}{V}\right)^{\frac{2}{3}} - 1\right]^3 B_0' + \left[\left(\frac{V_0}{V}\right)^{\frac{2}{3}} - 1\right]^2 \left[6 - 4\left(\frac{V_0}{V}\right)^{\frac{2}{3}}\right]\right\}$$

Where, $V_0$ is the equilibrium volume per unit cell, $B_0$ is the bulk modulus at zero pressure and $B_0' = \left(\frac{\partial B}{\partial P}\right)_T$ . This equation is been applied to our data for $Cu_2MnAl$ and $Hg_2CuTi$-type structure treating $V_0, E_0, B_0$ and $B_0'$ as fitting parameters. The optimized lattice parameter found from fig. 4.c. is 5.742Å, which is of similar value as obtained from earlier literatures [21][36][37][38].

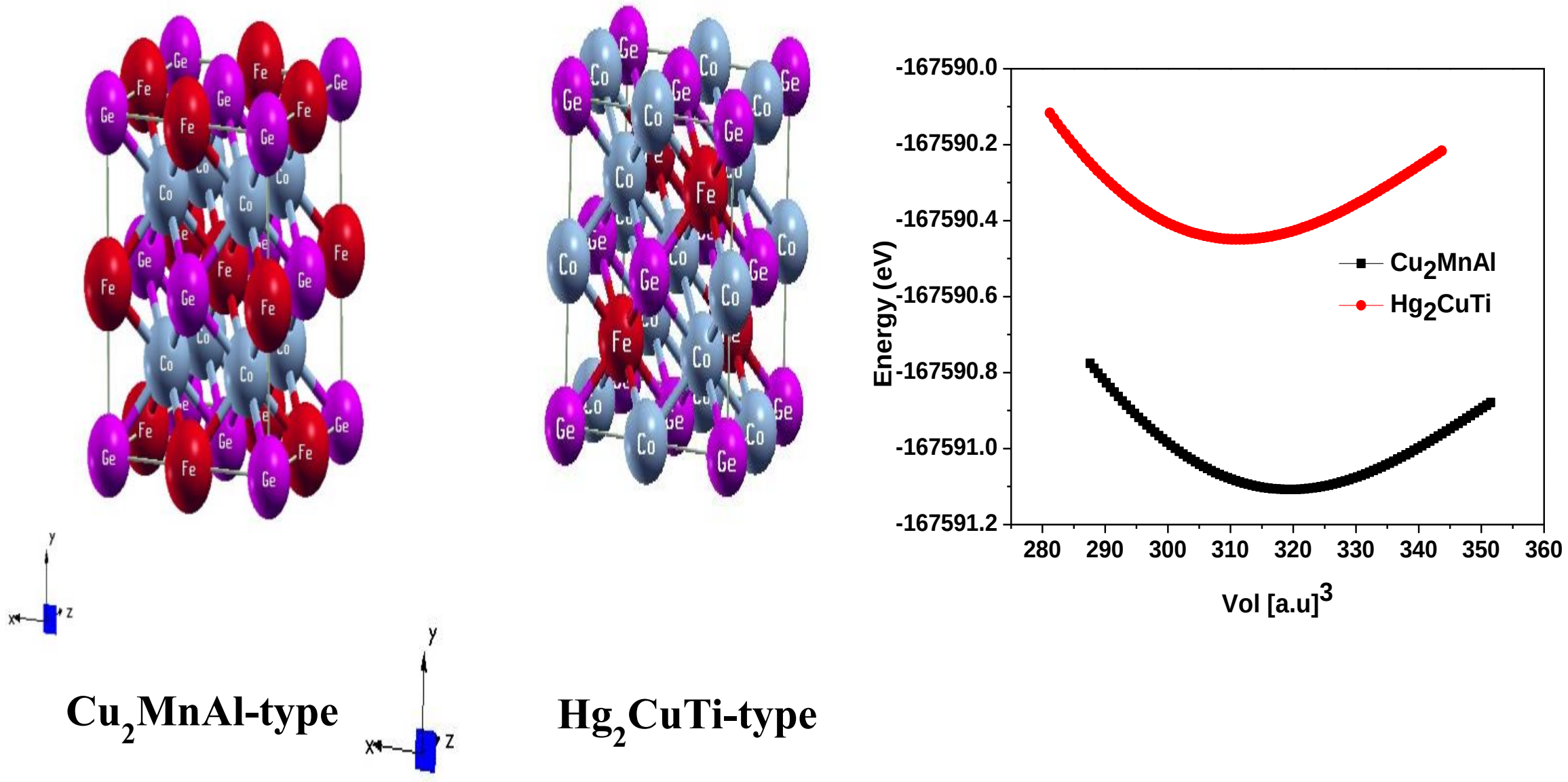

**a.** **b.** **c.**

**Fig.4.** Schematic representation of crystal structure of $Co_2FeGe$ Heusler alloy, **a.** $Cu_2MnAl$-type, **b.** $Hg_2CuTi$-type, **c.** Calculated total energy vs lattice volume of $Co_2FeGe$ Heusler alloy in $Cu_2MnAl$-type and $Hg_2CuTi$-type structure. Minimum of energy corresponds to equilibrium volume.

The chemical stability of the alloy was verified by means of formation energy, which is the excess energy of the bulk counterpart relative to the constituents. It can be calculated as:

$$E_{formation} = E^{tot}_{Co_2FeGe} - [\,2E^{bulk}_{Co} + E^{bulk}_{Fe} + E^{bulk}_{Ge}\,]$$

where, $E^{tot}_{Co_2FeGe}$ is the total energy of the bulk alloy compound at the equilibrium lattice constant and $E^{bulk}_{Co}$, $E^{bulk}_{Fe}$ and $E^{bulk}_{Ge}$ are the energy of each atom when they are crystallized in pure metal structure [17]. The values of the calculated formation energies in different structures is found in table 2. The negative value ensures the thermodynamic stability and the possibility of its experimental synthesis.

**Table 2.** Calculated lattice parameter, equilibrium energy and formation energy values of $Co_2FeGe$ HA in different structures.

| Structure | Space Group | Atomic Positions | a (Å) | $E_{min}$ (eV) | Formation energy (eV) |
|---|---|---|---|---|---|
| $Cu_2MnAl$-type | No: 225 Fm-3m | Co: 8c(1/4,1/4,1/4) and (3/4,3/4,3/4) Fe: 4b(1/2,1/2,1/2) Ge : 4a(0,0,0) [23] | 5.742 | -167591.121 | -1.349 |
| $Hg_2CuTi$-type | No: 216 F-43m | Co: 4a(1/4,1/4,1/4) and (1/2,1/2,1/2) Fe : (3/4,3/4,3/4) Ge : (0,0,0) | 5.692 | -167590.543 | -0.691 |

**3. c. Electronic and magnetic properties**

Comparison of our results with previous reports can be found from table 3. It is found that GGA method is unable to explain the theoretical Slater-Pauling (SP) moment value of $M_{tot}$

= ($N_v$ -24) = 6 $\mu_B$/f.u., (where no. of valence electrons, $N_v$ =2 Co ($3d^7 4s^2$)+ Fe ($3d^6 4s^2$)+ Ge ($4s^2 4p^2$) =30 [39] ) whereas, GGA+U method results more accurate SP-value of magnetic moment of the alloy , as this method takes into account the strongly correlated nature of the alloy and thus incorporates the local electron-electron correlation effect between the d-electrons of the transition metal atoms [23]. In our results, the on-site coulomb parameter (U) was assumed for the transition metals to be U (Co) = 3.9456 eV and U (Fe) = 3.8095 eV [36]. The addition of U- parameter to GGA method results higher moments for both Co and Fe atoms (table 3). It is seen that the moment of Fe atom is higher than that of the Co atom, as the Fe atom shows an complete exchange splitting of valence and conduction bands [23]. After U-value is incorporated, the change in moment becomes higher for Fe atom than Co atom, implying that the Fe moment interacts more strongly than Co moment [40]. Also, with GGA+U method the alloy seems to show perfect half-metallic character.

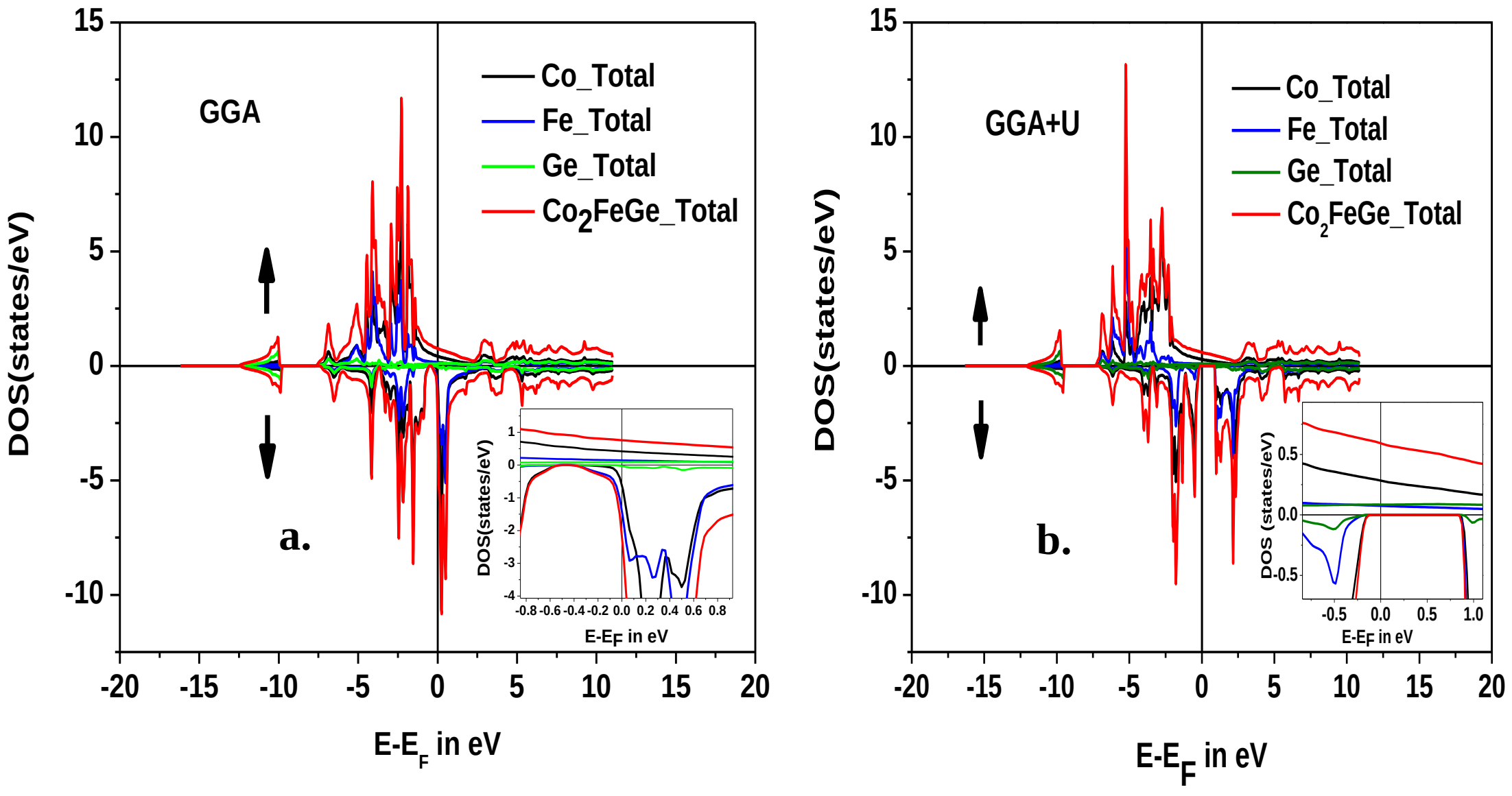

**Fig.5**. Calculated spin resolved density of states for $Co_2FeGe$ HA using **a.** GGA and **b.** GGA+U approximations.

**Table 3.** Calculated lattice parameter, magnetic moment and spin-polarization values of $Co_2FeGe$ HA in $L2_1$ structure

| Material | | Lattice parameter (Å) | | Magnetic moment ($\mu_B$/f.u.) | | Spin-polarisation |
|---|---|---|---|---|---|---|
| | | Theory | Expt. | Theory | Expt. | Theory |
| $Co_2FeGe$ | Previous Reports | 5.75 [38]<br>5.75 [39] | 5.736 [38]<br>5.702 [39]<br>5.738 [37] [41] | 5.61(GGA) [38]<br>5.693(GGA) [39]<br>5.999(GGA+U) [39]<br>5.72(GGA) [42]<br>6.02(GGA+U) [42]<br>5.758 (GGA) [36]<br>5.762(GGA+U) [36]<br>5.70 (GGA) [37]<br>5.442(LSDA) [19] | 5.74 [38]<br>5.54 [39]<br>5.9 [37] | 0.54(GGA) [39]<br>1(GGA+U) [39]<br>0.51[37]<br>1(LSDA+U) [36] |
| | This work | 5.742 | 5.72 | 5.69 (GGA)<br>$m_{Co}$ = 1.422 ;<br>$m_{Fe}$ = 2.919 ;<br>$m_{Ge}$ = 0.010 ; | 6.1 | 1 |
| | | | | 5.99 (GGA+U)<br>$m_{Co}$ = 1.549 ;<br>$m_{Fe}$ = 3.254 ;<br>$m_{Ge}$ = -0.046 ; | | |
| | | | | 6 (Slater-Pauling rule) | | |

It is evident from our results in table.2 that the major contribution towards the total magnetic moment of $Co_2FeGe$ comes from the Co and Fe atoms due to the hybridization between their 3d electrons [39] and Ge has no effective contribution to the total magnetic moment of the alloy, rather it has a role to give structural stability of the alloy [43].

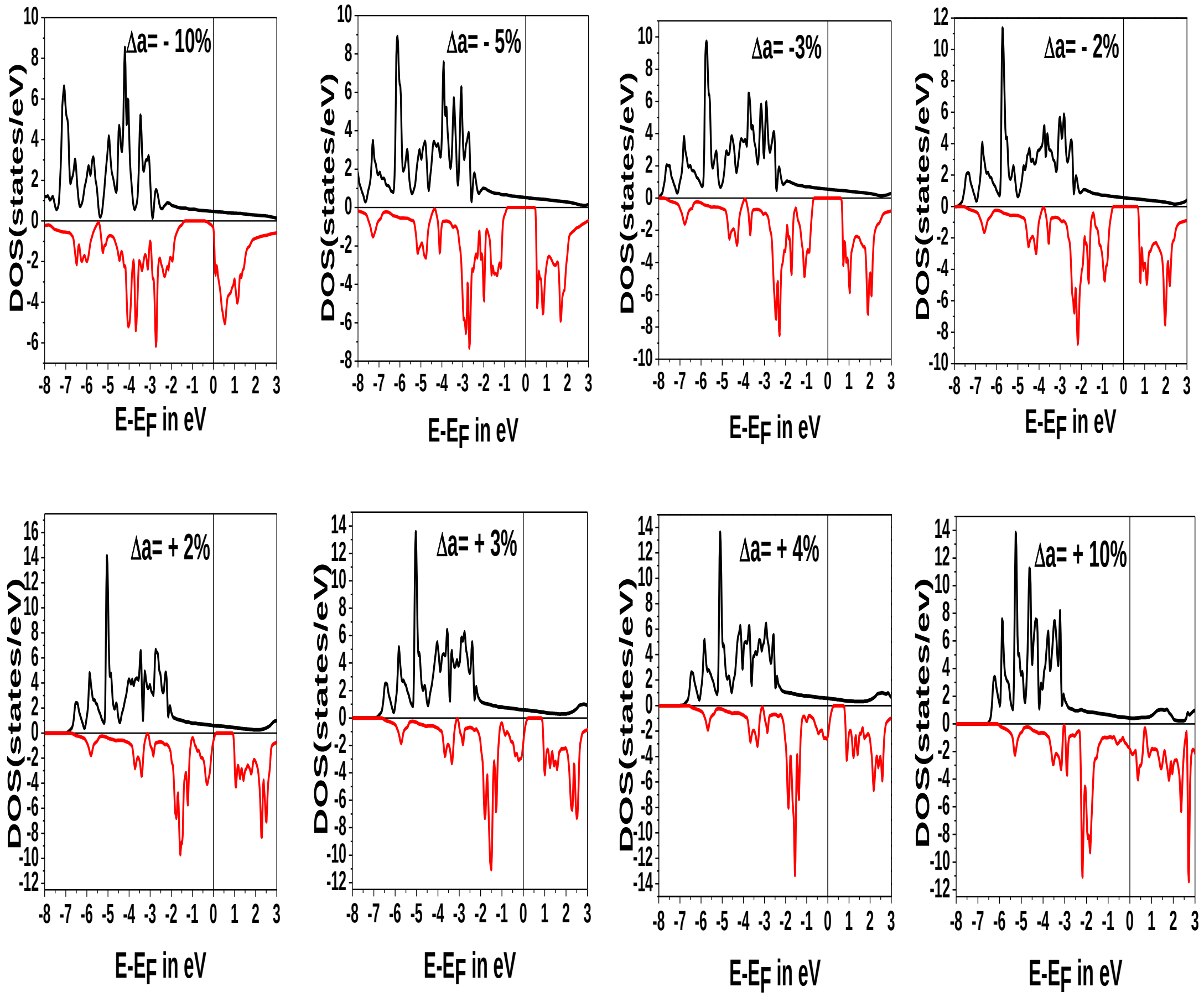

**Fig.6.** Effect of lattice strain on the spin-resolved density of states of $Co_2FeGe$ Heusler alloy.

Lattice strain (Δa) has a significant effect on the half-metallicity and the magnetic moment of the alloy. It is defined to be the change in lattice parameter with respect to its equilibrium value, $\Delta a = (a-a_0)/a_0$. Fig.6. shows how the density of states near the Fermi level changes with compressive (negative) and extensive (positive) strain. Half-metallic property holds for the range Δa = -5% to the equilibrium lattice constant ($a_0$), as shown in fig.7.b. With compression, the Co and Fe d-states hybridize more strongly resulting in increasing the minority gap; also the Fermi energy level moves towards the conduction band with compression [44]. Negative spin-polarization implies that the minority density of states at the Fermi level becomes higher than the majority states. Fig. 7.a. shows that the tensile strain increases the total moment of the alloy, where the increase in contribution of the moment mainly comes from Co atom.

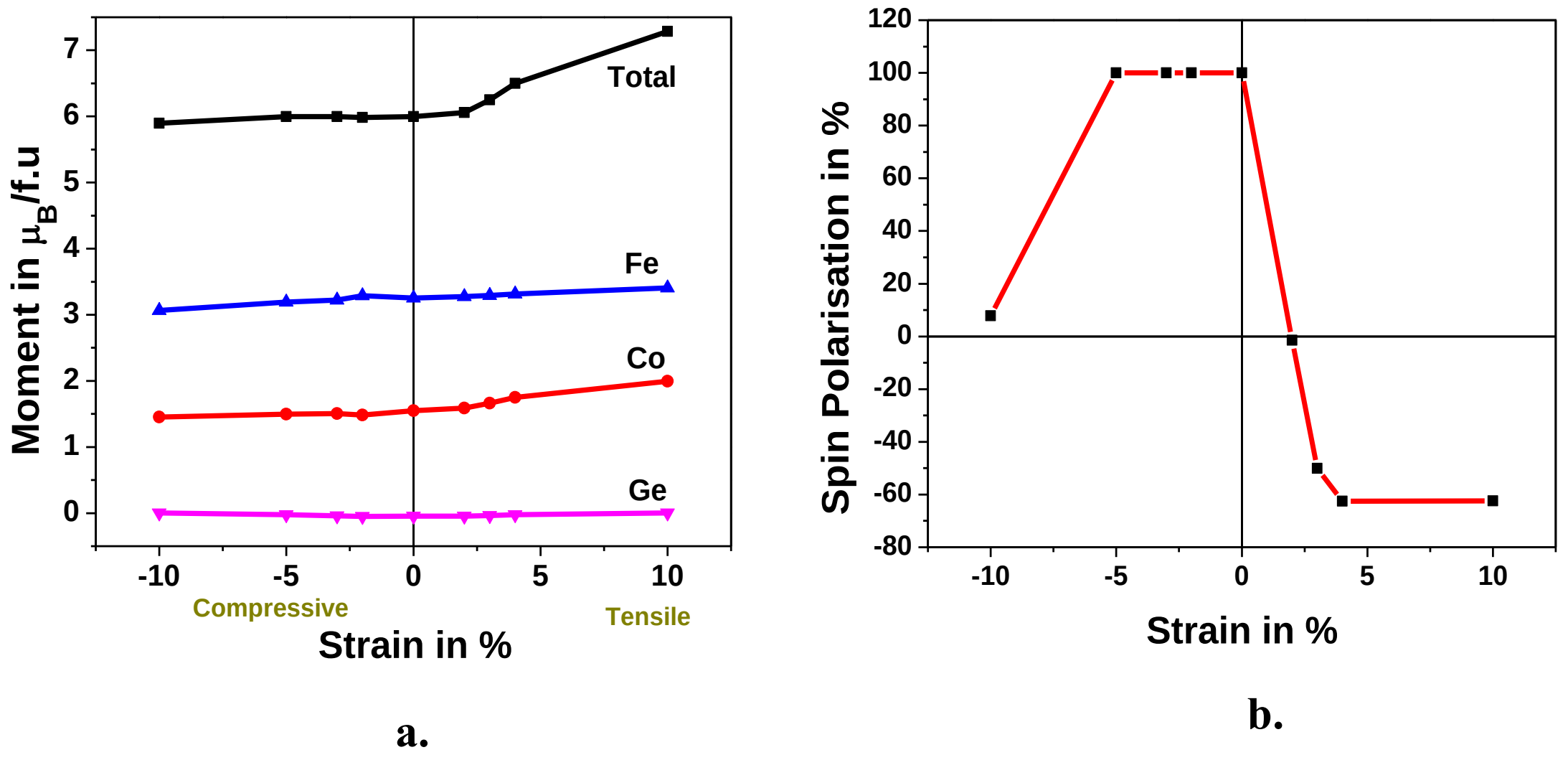

**Fig.7.** a. Effect of strain on total magnetic moment of $Co_2FeGe$ and individual atomic moment, b. Effect of strain on spin polarization of $Co_2FeGe$ Heusler alloy.

Generally Co-based full HAs exhibit localized moments and behave as strongly correlated materials [23][24] and the role of electron-electron correlation is important in this case [36]; Electron correlation strength depends on the on-site coulomb potential (U) and it influences the magnetic ordering in Heusler alloys [45]. It is also noticed in our computational result that GGA+U gives more accurate result than GGA only, when compared to the experimental results. Here, the strength of onsite electron-electron correlation is given by two parameters: the on-site coulomb part (U) and the on-site exchange part (J) such that $U_{eff}$ = U-J; in our calculations J was set to zero always and the value of U for Co and Fe were varied within compatible range and the effect was clearly seen on the total and individual magnetic moments of the $Co_2FeGe$ alloy and its transition metal atoms (Co and Fe). Figure 8. shows the variation of total magnetic moment of $Co_2FeGe$ alloy with the variation of values of $U_{eff}$-parameter for Co and Fe. We have varied individual atomic value of U in the range of 0-1 Ry (1Ry= 13.6056eV) keeping in mind the theoretical Slater-Pauling moment and our experimental result. The shaded gray plane is the theoretical Slater-Pauling moment value of $M_{tot}$ = 6 $\mu_B$/f.u. It is seen that the total moment increases with the U (Co) value for a constant value of U (Fe) and vice versa. We can see the range of values of U (Co) and U (Fe) for which the theoretical SP value is retained. The result is also important for explaining the deviated experimental value of moment (6.1 $\mu_B$/f.u.) from the theoretical one (6 $\mu_B$/f.u.).

## 4. Discussions

Balke et.al [1] had predicted the curie temperature of $Co_2FeGe$ HA which is having theoretical magnetic moment value of 6 $\mu_B$/f.u, to be greater than 1000K. The curie temperature of the cobalt based HAs are also found to obey a linear relationship with the magnetic moment: $T_c$ = 23+181*$M_{tot}$ [23][24], which is nearly satisfied by our experimental result ($T_c$=1073K). We find that the measured saturation magnetization at 0K is 6.1 $\mu_B$/f.u which very nearly matches with the theoretically predicted Slater-Pauling value of 6 $\mu_B$/f.u. This increase in moment is may be either due to the lattice strain present in the system (fig 7.a.) or the disorder induced increase in electron-electron correlation strength of the transition metal atoms present in the alloy. The alloy is having very high stiffness constant (D = 10.4 $nm^2$-meV) and exchange constant (A) values. The value of D is found by fitting the temperature dependent saturation magnetization result (fig.3) with equation (4) and thereafter A is calculated from equation (2).

We also find that the alloy remains half-metallic over a considerable range of lattice strain as evident from fig 6. and fig 7.b. It is seen from table.2 that with the inclusion of U, moment of both Co and Fe increases resulting increase in total moment. Increase in moment is higher for Fe atom, implying that Fe atoms interacts more strongly when effect of U is taken into account [46]. For a constant value of U (Fe), moment of Co atom increases with U (Co) and moment of Fe atom has a little decreasing trend, whereas for a constant value of U (Co), moment of Fe atom increases with U (Fe) and moment of Co atom stays nearly constant. We also explored the effect of electron correlation parameter (U) for Co and Fe on the spin polarization of the alloy (fig 9). It is seen that the alloy is 100% spin polarized for the range U(Co) = (0 - 0.2) Ry ((1Ry= 13.6056eV)) and U(Fe) = (0.2 - 0.8) Ry (1Ry= 13.6056eV) ; beyond this range the spin polarization drops as spin-dependent total density of states at the Fermi level ($E_F$) changes its values ( inset of fig 9). This optimization of results helps us to justify our experimental values of magnetic moment and stiffness constant (exchange constant).

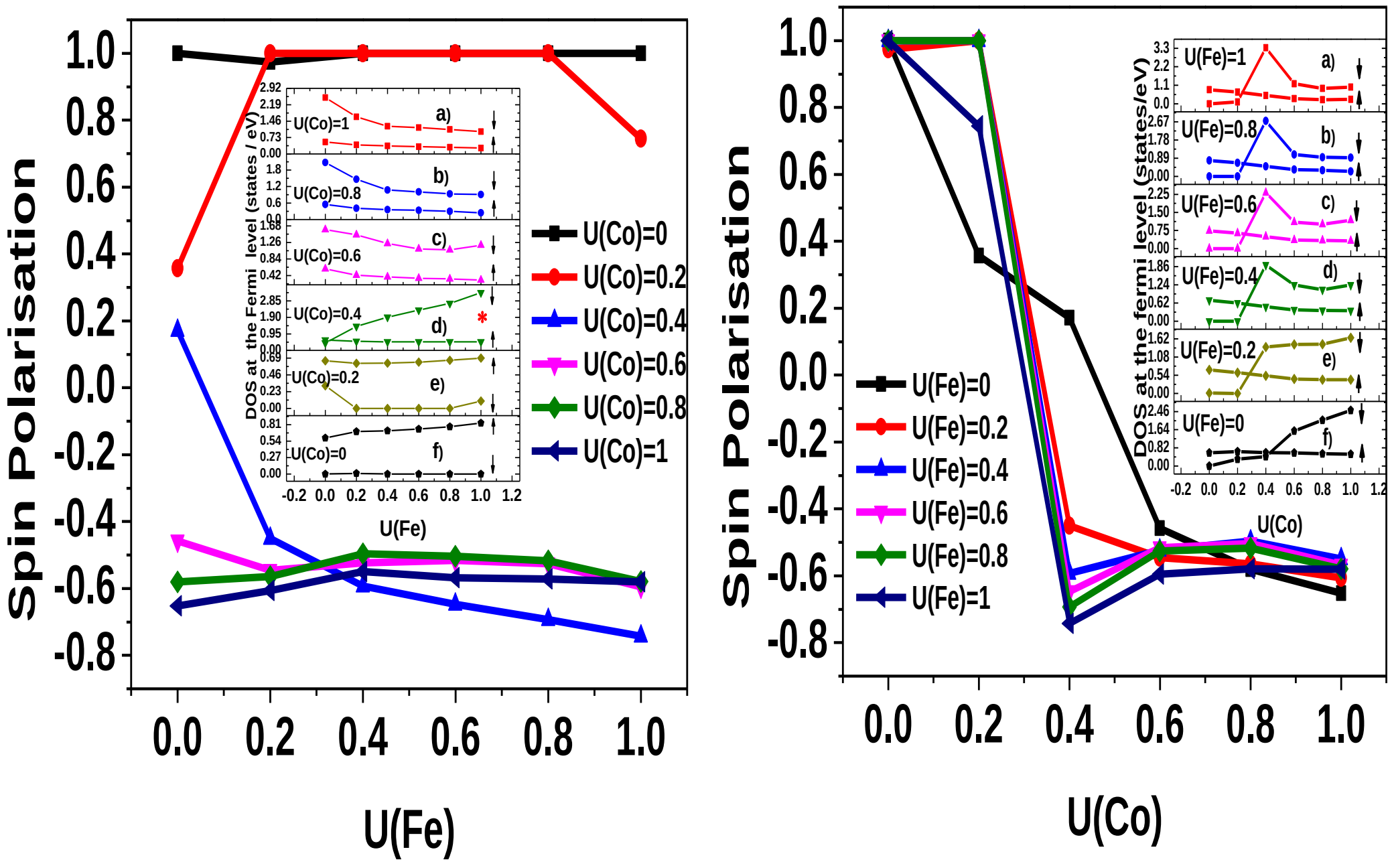

**Fig.9**. Dependence of spin-polarisation on electron-electron correlation parameter (U) of Co and Fe in $Co_2FeGe$ HA. ( inside image shows the variation of spin-dependent density of states at the Fermi level ($E_F$).

## 5. Conclusions

We have carried out a thorough experimental and computational studies on electronic and magnetic properties of $Co_2FeGe$ HA. It is shown that the alloy is having a very high magnetic moment and curie temperature and our experimental results satisfies the Slater-Pauling rule very closely. We have also studied the effect of strain on the half-metallicity of the alloy and found that it holds the spin-polarization up to a considerable amount against the equilibrium value. The stiffness constant of the alloy is also found to be very high. The experimental results are well justified by considering the electron-electron correlation effect in the alloy. Our study ensures that the alloy is an excellent candidate for spintronic applications.